\documentclass[12pt]{article}
\usepackage[dvips]{graphicx}

\begin{document}

\date{\today }
\title{Quantal Extension of Mean-Field Dynamics$^*$\footnote{$^*$
This work is supported in part by 
the U.S. DOE Grant No. DE-FG05-89ER40530.}}        
\author{\underline{D.~Lacroix}$^{a}$, Ph. Chomaz$^{a}$, S. Ayik$^{b}$ \\
{\small (a) {\it G.A.N.I.L., B.P. 5027, F-14021 Caen Cedex, France.}} \\
{\small (b) {\it Physics Department,Tennessee Technological University,
Cookeville TN38505, USA.}}}
\maketitle

\begin{abstract}
A method is presented for numerical implementation of the extended TDHF
theory in which two-body correlations beyond the mean-field approximation
are incorporated in the form of a quantal collision term. The method is
tested in a model problem in which the exact solution can be obtained
numerically. Whereas the usual TDHF fails to reproduce the long time
evolution, a very good agreement is found between the extended TDHF and the
exact solution.
\end{abstract}

\newpage

\section{Introduction}

The study of nuclei in out-off equilibrium configurations in many-body
quantum theory is a very complicated problem. However, due to their long
mean-free path inside the nucleus and to their large amplitude zero point
motion associated to a strong delocalization of their wave-packet, nucleons
could be often regarded as independent fermions moving within a mean
potential. Indeed, such theory, called Time-Dependent Hartree-Fock theory
(TDHF), first applied to nuclear dynamics some twenty years ago by Bonche 
{\it et al} \cite{Bon}, is able to reproduce quantitatively many properties
of Heavy-Ion collisions\cite{Nor}. However, in this theory, nucleons
interact only through the mean-field and collisions between particles are
neglected. It is difficult to believe that correlations induced by
collisions do not play also an important role when highly dissipative
processes are involved. For example, thermalization of single-particle
degrees of freedom\cite{Cho1} and damping of collective motion\cite
{Giant,Diss} could only be understood by the introduction of two-body
correction. Furthermore, usual mean-field theory is not able to reproduce
the large experimental width of mass produced during Heavy-Ion reactions at
Intermediate energies.

In the extended Time-Dependent Hartree-Fock (ETDHF) theory, the description
is improved beyond the mean-field approximation by incorporating
correlations in the form of a quantal collision term\cite{ETDHF}. Due to the
numerical complexity, the applications of this theory on realistic
situations remain a difficult problem, and therefore only a few approximate
calculations have been performed so far\cite{app_etdhf}. In this paper, we
describe the basic features of the ETDHF approach, propose a method for
obtaining numerical solutions of the theory, and present an application of
the method to an exactly solvable model.

\section{Extended Time-Dependent Hartree-Fock Theory}

The exact description of a quantum system is contained in the many-body
density operator $\hat{D}(t)$ and its dynamical evolution is given by the
Liouville-von Neumann equation 
\begin{eqnarray}
i\hbar \frac{d\hat{D}}{dt}=\left[ \hat{H},\hat{D}\right]   \label{Neumann}
\end{eqnarray}
Since the system has a {\it a priori} a large number of degrees of freedom,
the practical resolution of equation (\ref{Neumann}) is often impossible.
The basic assumption of mean-field theory is to suppose that only one-body
observables are important in the description of the system (neglecting , by
the way, all correlations of order greater or equal to two). Usual
mean-field theory neglect completely two-body dynamics and values of
two-body observables are obtained assuming that they are equal to the least
biased value when one-body observables are known\cite{Balian}. In this
approaches, all relevant information is contained in the one-body density
operator $\hat{\rho}$ and the dynamical evolution is replaced by\cite{Ring} 
\begin{eqnarray}
i\hbar \frac{d\hat{\rho}}{dt}=\left[ \hat{h},\hat{\rho}\right] 
\end{eqnarray}
where $\hat{h}$ is the mean field associated to $\hat{H}$. This approximate
theory is already a many-body theory for out-off equilibrium finite system.
Note that, even if two-body dynamics are not explicitly included, non
trivial correlations between single particle states exists due to the
reorganization of the mean-field during the time evolution. \\ However, this
two-body correlation is not sufficient to reproduce the many facet of nuclei
desexcitation. In particular, no dissipation, characteristic of the
irreversible flow from collective motion to single particle degrees of
freedom\cite{Cho1} and expected to be responsible of the thermalization of
nuclei, is included. The next step towards extended mean-field is to include
information about two-body correlation dynamics. This could be obtained in
particular by a truncation of the BBGKY hierachy\cite{BBGKY} to the two
first equations. This approaches assumed that TDHF is already a good
approximation of the dynamics and two-body correlations are supposed only to
act as a small perturbation added on top of the mean-field. This leads to a
correction factor in the evolution of the one-body density operator\footnote{%
In this expression, usually, an additional term exist due to the propagation
of initial correlations. In this paper, we suppose that, initially, we have
uncorrelated states.}\cite{Abe} 
\begin{eqnarray}
i\hbar \frac{\partial \hat{\rho}}{\partial t}=\left[ \hat{h},\hat{\rho}%
\right] +K(\hat{\rho})  \label{evol}
\end{eqnarray}
with 
\begin{eqnarray}
\begin{array}{cc}
K(\hat{\rho})=-\frac i\hbar \int\limits_{-\infty }^t{dt^{\prime }}%
tr_2[V_{12}, & U_{12}\left( t,t^{\prime }\right) \rho _1\rho _2\widetilde{%
V_{12}}\left( 1-\rho _1\right) \left( 1-\rho _2\right) U_{12}\left(
t^{\prime },t\right)  \\ 
& -U_{12}\left( t,t^{\prime }\right) \left( 1-\rho _1\right) \left( 1-\rho
_2\right) \widetilde{V_{12}}\rho _1\rho _2U_{12}\left( t^{\prime },t\right) ]
\\ 
\label{ETDHF} & 
\end{array}
\end{eqnarray}
Taken same notations as in ref.\cite{Abe}, the label $1$ and $2$ refers to
first and second particles, for example 
\begin{eqnarray}
\left\langle ij\left| \rho _1\rho _2\right| kl\right\rangle =\left\langle
i\left| \hat{\rho}\right| k\right\rangle \left\langle j\left| \hat{\rho}%
\right| l\right\rangle 
\end{eqnarray}
$U_{12}$ is the propagator of two independent particles associated to the
mean-field: 
\begin{eqnarray}
U_{12}=U_1\otimes U_2
\end{eqnarray}
with 
\begin{eqnarray}
U_1(t,t^{\prime })=T\left( \exp \left( -\frac i\hbar \int_{t^{\prime
}}^th(\rho \left( s\right) )ds\right) \right) 
\end{eqnarray}
and $tr_2$ is the partial trace taken over the second particle whereas $%
\widetilde{V_{12}}$ includes the antisymmetrization.

Eq. (\ref{ETDHF}) is the starting point of most of the theories that goes
beyond TDHF. For example, all semi-classical applications which include
collision effects\cite{BUU} (namely BUU approaches) can be derived from it,
by developing the Wigner transform of (\ref{evol}) up to first order in $%
\hbar $. Over the past decade, significant progresses have been made on the
description of nuclear dynamics within semi-classical transport theories and
extension of mean-field seems to be a promising tool for describing
dynamics. However, nuclei are quantum objects even under extreme conditions
and important physical component are missing in a semi-classical treatment.
Some attempts in order to treat the collision term in a quantum picture
already exists\cite{app_etdhf,Tohayama}. Nevertheless, due to the
complicated expression of $K(\rho )$, theses investigations have been
carried out under strong approximations.

Here, we describe a procedure for treating the collision term (\ref{ETDHF})
with minimum bias and which may be adapted for solving the ETDHF equation in
realistic situations under reasonable numerical approximations.

\section{Re-examination and numerical algorithm}

In a single particle bases, the one-body density operator reads 
\begin{eqnarray}
\hat{\rho} = \sum_{\lambda,\lambda^{\prime}}
\left|\lambda\right>n_{\lambda,\lambda^{\prime}}\left<\lambda^{\prime}\right|
\end{eqnarray}
and the evolution equation take the form\cite{Tohayama} 
\begin{eqnarray}
\left\{ 
\begin{array}{c}
i\hbar \frac{\partial}{\partial t}\left|\lambda\right> = h(\rho)
\left|\lambda\right> \\ 
\\ 
\frac{d}{dt}n_{\lambda,\lambda^{\prime}}=-\frac{1}{\hbar^2}\left(F_{\lambda,
\lambda^{\prime}} +F^*_{\lambda^{\prime},\lambda} \right)
\end{array}
\right.  \label{F_l}
\end{eqnarray}
with

\begin{eqnarray}
F_{\lambda,\lambda^{\prime}}=&&\sum_{\alpha,\alpha^{\prime},\beta,\beta^{%
\prime},\gamma,\delta, \delta^{\prime}} {\
\left<\lambda\delta^{\prime}|V_{12}|\alpha\beta\right>} \int_{-\infty}^{t} {%
\ dt^{\prime}} \left.\left<\alpha^{\prime}\beta^{\prime}|\widetilde{V_{12}}
|\gamma\delta\right>\right|_{t^{\prime}}  \nonumber \\
&& ( n_{\gamma\lambda^{\prime}} n_{\delta\delta^{\prime}}
(\delta_{\alpha\alpha^{\prime}}
-n_{\alpha\alpha^{\prime}})(\delta_{\beta\beta^{\prime}}
-n_{\beta\beta^{\prime}}) \\
&&-n_{\alpha\alpha^{\prime}} n_{\beta\beta^{\prime}}
(\delta_{\gamma\lambda^{\prime}}
-n_{\gamma\lambda^{\prime}})(\delta_{\delta\delta^{\prime}}-n_{\delta%
\delta^{\prime}}) )  \label{F__}
\end{eqnarray}

This expression is the most general one could extract from (\ref{ETDHF}). We
know from work of \cite{Tohayama}, that approximations in (\ref{F__}) leads
to important change in physical results. In particular, the time integration
is often replaced by a conservation of energy based on a Markovian
approximation of the collision but we know that this memory effect is
important since it creates the non-trivial coupling between collective and
single particle motions. Furthermore, we are obligated to consider the
non-diagonal part of (\ref{F__}) which will give the coherent evolution of
single-particle states.

A direct resolution of (\ref{F__}) needs a large numerical effort. However,
in the following, we will see that it could be considerably simplified
without loosing its generality.

\subsection{Coarse-Graining in time}

Supposing that we know the one-body density at time $t$: 
\begin{eqnarray}
\hat{\rho}=\sum_i{\left| \Phi _i\right\rangle n_i\left\langle \Phi _i\right| 
}
\end{eqnarray}

\begin{figure}[tbph]
\begin{center}
\includegraphics*[height=8cm,width=16cm]{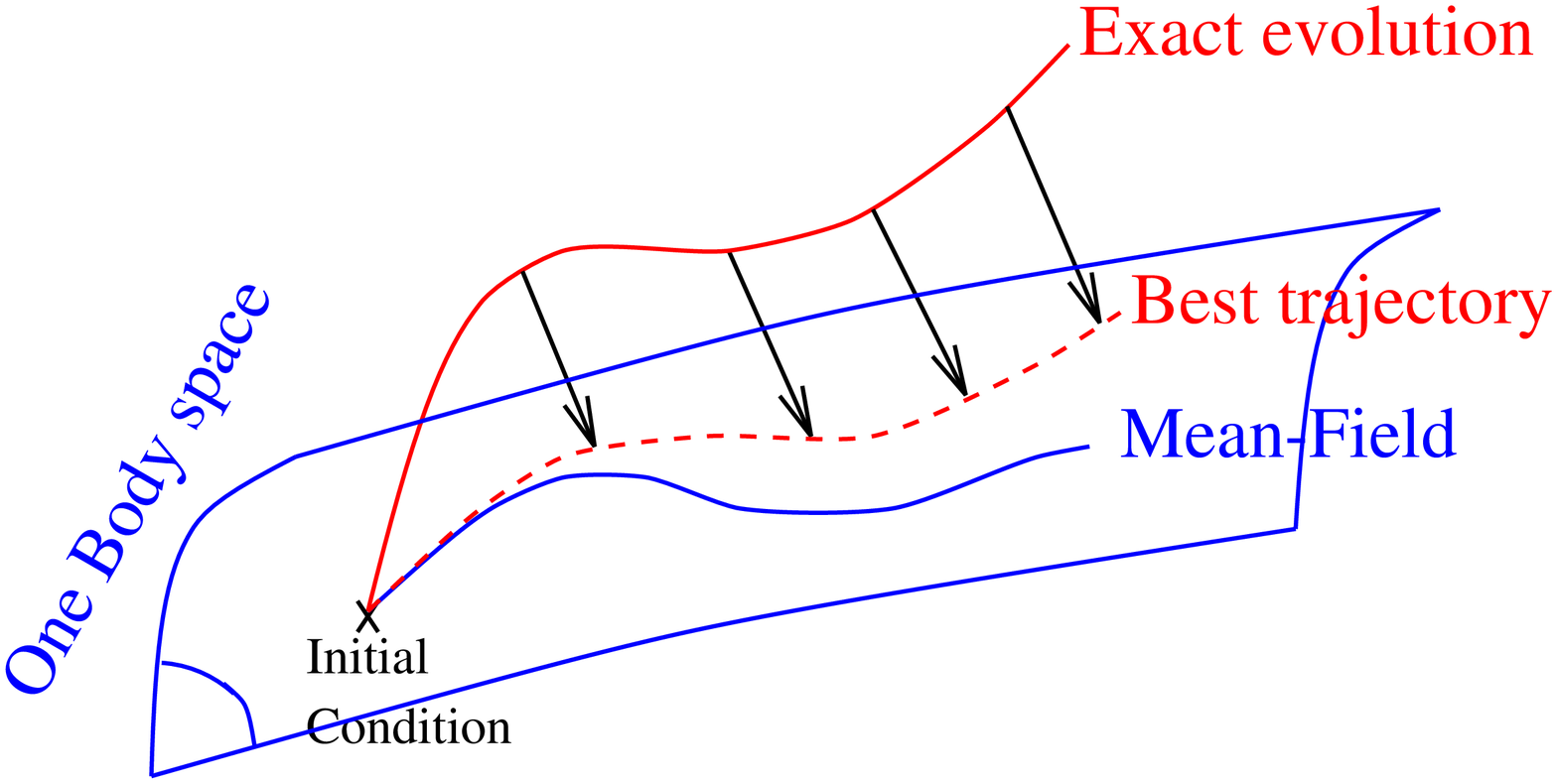}
\end{center}
\caption{Geometric picture of TDHF and EXACT evolution. The Initial
condition is supposed to be contained in the one-body space. The exact
trajectory is symbolised by a path in the total space. In particular, due to
the presence of two-body correlations, this trajectory do not remain in the
one-body manifold. The best associated trajectory in the one-body space is
represented in dashed-line whereas the TDHF evolution is represented in
solid line. The neglected two-body correlation dynamics in TDHF, implies
that this theory do not corresponds to the best trajectory for long time
evolution.}
\label{fig:1}
\end{figure}

From information theory, it means that we have access to the least biased
description of our system when only one-body observables are known. We can
have a geometrical picture of the information reduction. One-body space
could be represented by a manifold in a bigger space (see fig. \ref{fig:1}).
Suppose that we start from one point on this manifold. The exact dynamical
evolution could be represented schematically by a trajectory in this space.
Due to the presence of two-body correlations, the trajectory will not remain
in the one-body space. Nevertheless, the exact trajectory is associated to a
path in the one-body space which reproduce the one-body dynamics of the
exact evolution. The goal of using mean-field theory is to predict this
trajectory. In the one-body space, the TDHF theory corresponds also to a
trajectory. However, due to the neglected two-body correlations dynamics,
usual mean-field is not able to reproduce the long-time evolution of the
system. This is symbolized in fig. \ref{fig:1}, by a strong departure of the
TDHF theory from the best one-body trajectory. The goal of ETDHF is to
correct the mean-field evolution in order to be a much better approximation
of the best trajectory.

However, some remarks are in order:

\begin{itemize}
\item  Whereas TDHF fails to reproduce the long time evolution due to
accumulated errors on two-body observables, one-body dynamics remains
dominant. It implies that it exists a typical macroscopic time interval $%
\Delta t$ during which mean-field is a good approximation of the dynamics.
This time interval is of the order of the time between two collisions. Note
that, a basic assumption of ETDHF and TDHF is that this time interval is
large compared to the typical time-scale of mean-field evolution.

\item  For time greater or equal to $\Delta t$, two-body correlations appear
as a small correction to the mean-field and could be treated in perturbation.
\end{itemize}

\begin{figure}[tbph]
\begin{center}
\includegraphics*[height=6cm,width=10cm]{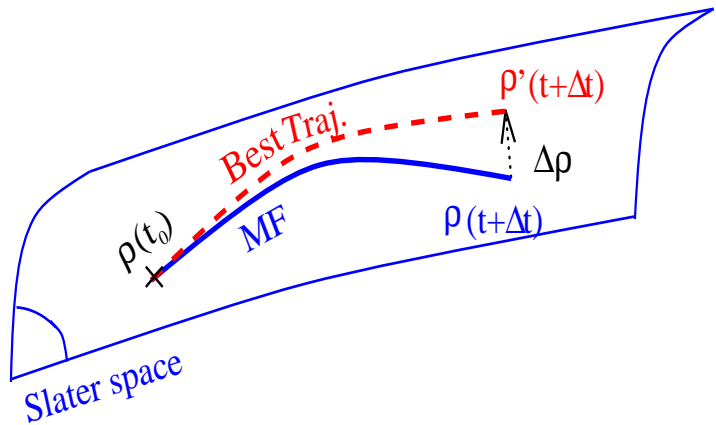}
\end{center}
\caption{Schematic representation of ETDHF evolution within the
coarse-graining time framework. The two steps procedure is represented. In
solid line, the usual TDHF is first performed. After a time $\Delta t$, a
correction is applied (small dashed line) in order to account for two-body
correlations propagation.}
\label{fig:2}
\end{figure}

\begin{figure}[tbph]
\begin{center}
\includegraphics*[height=8cm,width=14cm]{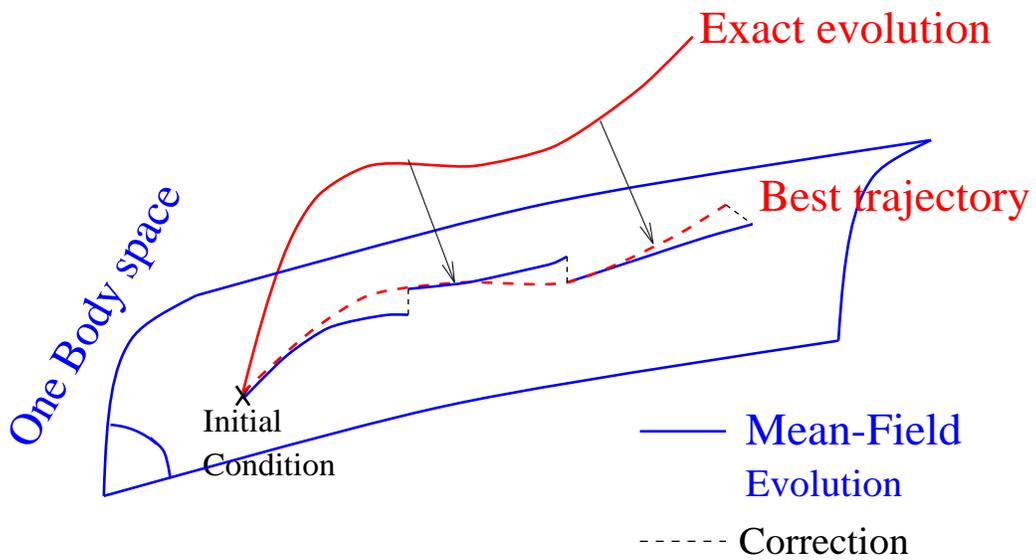}
\end{center}
\caption{With the same convention as in fig.1, the schematic evolution
through the numerical resolution of ETDHF is represented. Each $\Delta t$, a
correction is applied to the mean-field evolution. The ETDHF procedure is
expected to follow the best trajectory in the one-body manifold.}
\label{fig:3}
\end{figure}

It enable us to coarse-grain the time evolution for the extended mean-field
dynamics. Instead of solving equation (\ref{evol}) in time, we will divide
the evolution into two step:

\begin{itemize}
\item  The Hartree-Fock states will be evolved through the usual TDHF
equation between $t$ and $t+\Delta t$ 
\begin{eqnarray}
\left\{ 
\begin{array}{c}
i\hbar \frac \partial {\partial t}\left| \Phi _i\right\rangle =h(\rho
)\left| \Phi _i\right\rangle  \\ 
\\ 
\frac d{dt}n_i=0
\end{array}
\right.   \label{TDHF_1}
\end{eqnarray}

\item  We then use the perturbation theory in order to correct the one-body
density by the error accumulated during $\Delta t$ associated with (\ref{F_l}%
). We then find the new density operator $\hat{\rho ^{\prime }}$(see fig.\ref
{fig:2}) 
\begin{eqnarray}
\begin{array}{ccccc}
\hat{\rho ^{\prime }}(t+\Delta t) & = & \hat{\rho}(t+\Delta t) & + & 
\widehat{\Delta \rho } \\ 
&  & \Uparrow  &  & \Uparrow  \\ 
&  & i\hbar \frac{\partial \hat{\rho}}{\partial t}=\left[ h,\hat{\rho}%
\right]  &  & Integrated \\ 
&  &  &  & Effect \\ 
&  &  &  & of~collision
\end{array}
\label{delta_rho}
\end{eqnarray}
The new one-body density is then diagonalize and the procedure is iterated
(see fig. \ref{fig:3}).
\end{itemize}

\subsection{Determination of a Basis}

In order to express different operator matrices elements, we need {\it a
priori} a complete basis of the single-particle space. However, this could
rarely be obtained in a problem with a huge number of degrees of freedom. On
the other hand, if we suppose to have such a basis, only few of the states
will be necessary in order to have the main information about the system
(the rest corresponding to configuration which are not accessible). In a
previous application\cite{Tohayama}, the author proposes to include only a
part of the Hartree-Fock bases (including all occupied states (''hole
states'') and few unoccupied states (''particle states'')) and to follow
them with time. However, the system could reach many different
configurations and single-particle states that were not included at the
initial time could become important. It is thus difficult to believe that
one can choose the basis at the initial time. Conversely, at each
coarse-grained time step $\Delta t$, we will construct a bases that contains
the important information about different configurations accessible to the
system between a given time $t$ and $t+\Delta t$. According to this
procedure,

\begin{itemize}
\item  At a given time $t$, we suppose to have the Hartree-Fock basis $%
\left| \Phi _i(t)\right\rangle $. We only perform the evolution of occupied
states between $t$ and $t+\Delta t$.

\item  At time $t+\Delta t$, we complete the hole states ( $\left| \Phi
_i(t+\Delta t)\right\rangle $) by a limited ensemble of particle states( $%
\left| p_j(t+\Delta t)\right\rangle $). This states are constructed in order
to be a good approximation of eigenstates of $h(t+\Delta t)$ with an energy
smaller than a given energy $\varepsilon _{Max}$. In the following, we will
discussed in detail the construction of the basis and the choice of $%
\varepsilon _{Max}$.

\begin{description}
\item  $\Rightarrow $ Between $t$ and $t+\Delta t$, two single-particle
states (1-2) could collide. As a consequences, new states could be
populated. In a simplified vision of collision\footnote{%
Here, the simplification is contained in the fact that, in general, states
(1-2) are not eigenstates of $h$ which is the condition of having an exact
conservation of energy.}, we can assume that the transition probability from
(1-2) to two new states (3-4) is important when 
\[
\varepsilon _1+\varepsilon _2\sim \varepsilon _3+\varepsilon _4
\]
The maximum accessible energy $\varepsilon _{Max}$ is obtained when 1 and 2
are the hole state with highest energy $E_{Max}$ and 3 or 4 is the hole
state with lowest energy $E_{Min}$. This give the natural truncation energy
for the basis: 
\[
\varepsilon _{Max}=2E_{Max}-E_{Min}
\]

\item  $\Rightarrow $ In order to find states with low energy, we apply an
imaginary time method to Hartree-Fock states. This method, explained in
APPENDIX A, provides finally a truncated basis: $\displaystyle \left\{
\left| \Phi _i(t+\Delta t)\right\rangle \otimes \left| p_j(t+\Delta
t)\right\rangle \right\} $ where all different matrices could be expressed.
This basis will be called ''{\it instantaneous basis}'' and noted
generically $\left\{ \left| \lambda \right\rangle \right\} $ in the
following.
\end{description}

\item  In order to include the memory effect in the collision terms. The
instantaneous basis is evoluted backward self-consistently.
\end{itemize}

\section{Re-organization of occupation numbers and single-particle
wave-functions.}

The collision term introduce a small correction to the density at each
macroscopic time step $\Delta t$, see expression (\ref{delta_rho}). In the
following, we will treat $\widehat{\Delta \rho}$ in first order perturbation
theory. In this case, the new density operator could be expressed as: 
\begin{eqnarray}
\hat{\rho^{\prime}} = \sum_{i} \left|\Phi^{\prime}_\lambda\right>
n^{\prime}_\lambda\left<\Phi^{\prime}_\lambda\right|
\end{eqnarray}
where new states are obtained through the following procedure

\begin{itemize}
\item  {\bf New occupation numbers:} In first order perturbation, only
diagonal elements of $\widehat{\Delta \rho }$ are necessary to calculate the
new eigenvalues $n_\lambda ^{\prime }$ of $\hat{\rho ^{\prime }}$. The time
evolution of this diagonal elements between $t$ and $t+\Delta t$ is given by
a {\it master equation} (see APPENDIX B)

\begin{eqnarray}
\frac{dn_\lambda }{dt}=(1-n_\lambda ){{\cal W}_\lambda ^{+}}-n_\lambda {%
{\cal W}_\lambda ^{-}}  \label{master}
\end{eqnarray}
with initial condition 
\begin{eqnarray}
n_\lambda ^{\prime }(t)=n_\lambda (t)
\end{eqnarray}
During $t$ and $t+\Delta t$ the gain and loss term, respectively ${\cal W}%
_\lambda ^{+}$ and ${\cal W}_\lambda ^{-}$, are considered constant. In this
case, the equation (\ref{master}) is exactly solvable and reads: 
\begin{eqnarray}
n_{\lambda ^{\prime }}(t+\Delta t) &=&n_\lambda (t)\exp \left( -\Delta
t\left( {\cal W}_\lambda ^{+}+{\cal W}_\lambda ^{-}\right) \right)  
\nonumber \\
&&+\frac{{\cal W}_\lambda ^{+}}{{\cal W}_\lambda ^{+}+{\cal W}_\lambda ^{-}}%
\left( 1-\exp \left( -\Delta t\left( {\cal W}_\lambda ^{+}+{\cal W}_\lambda
^{-}\right) \right) \right) 
\end{eqnarray}

\item  {\bf Reorganization of states:} The new states are given in
perturbation theory by 
\begin{eqnarray}
\left| \phi _\lambda ^{\prime }(t+\Delta t)\right\rangle =\left| \lambda
\right\rangle +\sum_{\lambda ^{\prime }\neq \lambda }{\frac 1{n_\lambda
-n_\lambda ^{\prime }}\left| \lambda ^{\prime }\right\rangle \left\langle
\lambda ^{\prime }\right| \Delta \rho \left| \lambda \right\rangle }
\end{eqnarray}
It is important to note, that, not only previously occupied states but also
part of unoccupied states are included at each macroscopic time-step in the
dynamics. In a previous application\cite{Tohayama}, Tohayama argued that
non-diagonal elements in the density matrices are important. In our
approach, they are explicitly included since the first order perturbation is
equivalent to a approximate diagonalization of $\hat{\rho}^{\prime }$.
However, here we do not fixe the basis at the initial time and the system
will dynamically choose which configuration will be accessible.
\end{itemize}

In this section, we have described the scheme we will use in order to extend
TDHF. In the following, we will apply it to a model that could be exactly
solved numerically.

\section{Application}

We have considered two distinguishable particles in a one-dimensional
anharmonic potential. The total Hamiltonian reads 
\begin{eqnarray}
H=\sum_i{\left( \frac{{\hat{p}_i}^2}{2m}+U_i\right) }+\sum_{i<j}{V_{ij}}
\end{eqnarray}
where the one-body external field is 
\begin{eqnarray}
U_i=\frac 12k{\hat{x}_i}^2+\frac 14k^{\prime }{\hat{x}_i}^4
\end{eqnarray}
and the non-local two-body interaction $V_{ij}$ is taken as 
\begin{eqnarray}
V_{i,j}=\frac{t_0}{\sqrt{2\pi }\sigma }{\exp -\frac{\left( \hat{r}_i-\hat{r}%
_j\right) ^2}{2\sigma ^2}}
\end{eqnarray}
This problem, which seems simple in appearance contains already different
features of many-body physics and is a good benchmark for ETDHF since it
could be compared with the exact solution.

\subsection{Initial conditions and evolution}

\begin{figure}[tbph]
\begin{center}
\includegraphics*[height=8cm,width=8cm]{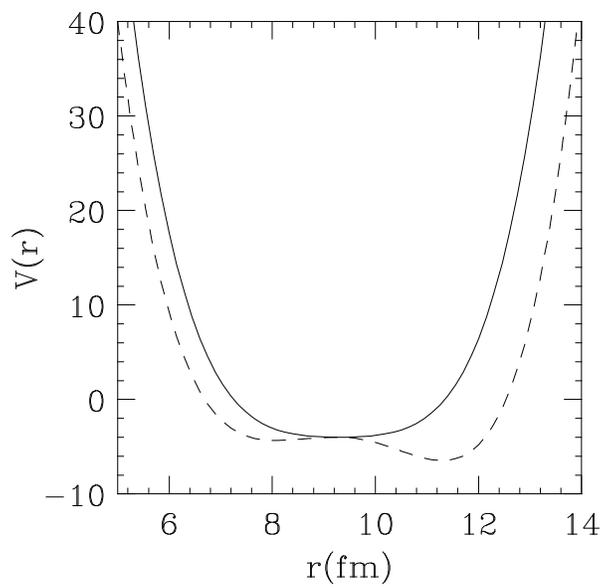}
\end{center}
\caption{One-body part of the external field in r-space, with the initial
constraint (dashed line) and without the initial constraint (solid line).}
\label{fig:4}
\end{figure}

We consider the system initially heated and constrained. The initial
two-body density operator is written as a statistical equilibrium 
\begin{eqnarray}
\hat{D}^{ini}=\frac 1{{\cal Z}}\exp -\frac 1{k_BT}\left( \hat{H}-\lambda 
\hat{Q}\right)
\end{eqnarray}
where $\hat{Q}$ is a one-body operator and $\lambda \hat{Q}$ is the initial
constraining field. At initial time, the constraint is relaxed and the
system evolve. In fig. \ref{fig:4}, we show the one-body external field part
of $\hat{H}$ in r-space (solid line) and the equivalent constrained field $%
\hat{H}-\lambda \hat{Q}$(dashed line). After relaxation of the constraint,
three different evolutions are compared:

\begin{itemize}
\item  {\bf EXACT evolution:} For the exact evolution, we solve exactly the
von Neumann equation for the two-body density operator: 
\begin{eqnarray}
i\hbar \frac{d\hat{D}}{dt}=\left[ \hat{H},\hat{D}\right] 
\end{eqnarray}

\item  {\bf TDHF:} The equation of motion is 
\begin{eqnarray}
i\hbar \frac{\partial \rho }{\partial t}=\left[ h,\rho \right] 
\end{eqnarray}
Note that occupation numbers remains unchanged during the evolution 
\begin{eqnarray}
\frac{dn_i}{dt}=0
\end{eqnarray}

\item  {\bf ETDHF:} For the ETDHF, we apply the procedure described above.
In this case, occupation numbers evolve in time and new states are mixed to
initial one during the dynamics.
\end{itemize}

For both TDHF and ETDHF, the initial density is the one-body that
corresponds to the exact one 
\begin{eqnarray}
\hat{\rho}^{ini}=tr_2\hat{D}^{ini}
\end{eqnarray}

\subsection{Results and Discussion}

\subsubsection{Prediction of one-body dynamics:}

\begin{figure}[tbph]
\begin{center}
\includegraphics*[height=16cm,width=13cm]{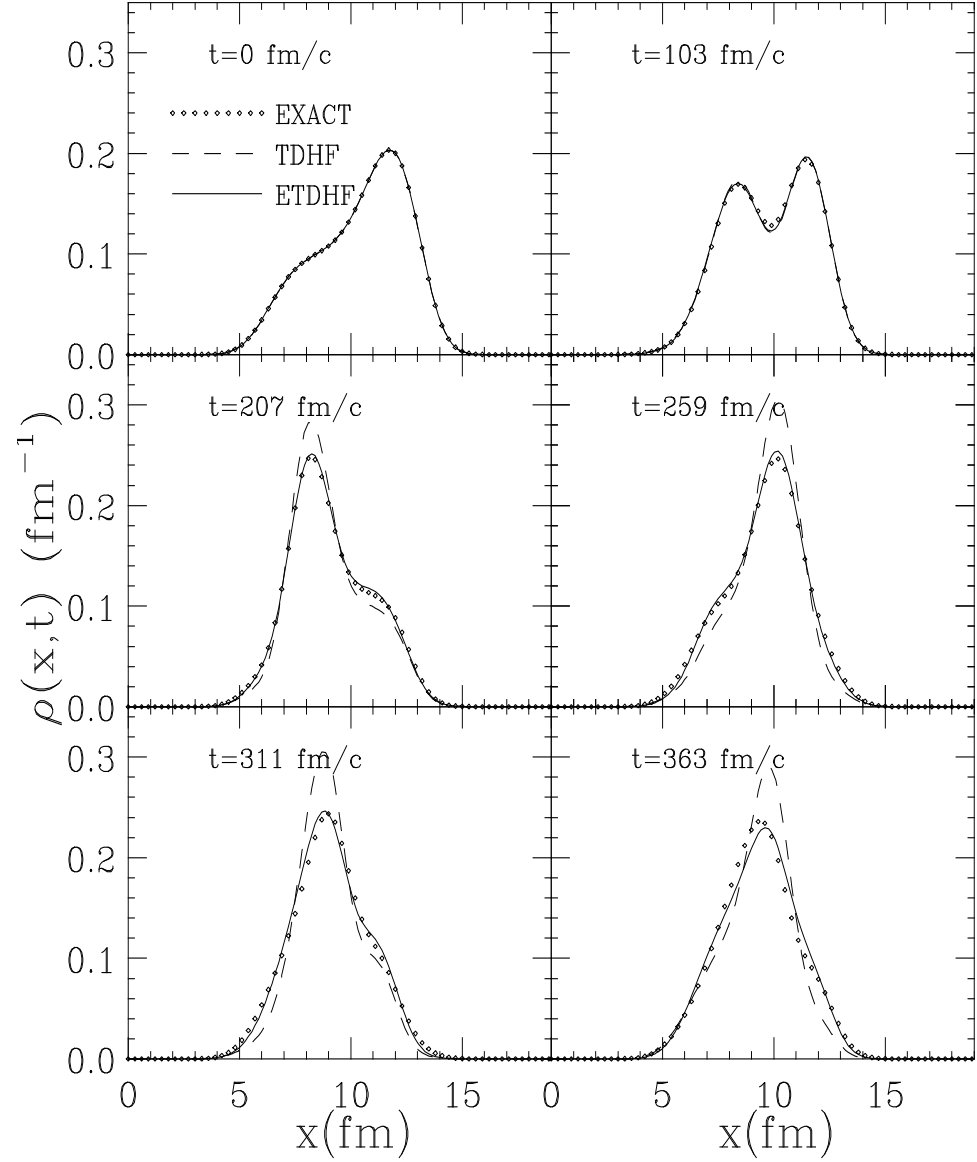}
\end{center}
\caption{Evolution of the one-body density operator in r-space. Three
different simulations are displayed: Exact (small circle), TDHF (dashed
line) and ETDHF (solid line).}
\label{fig:5}
\end{figure}

In fig. \ref{fig:5}, we have represented the dynamical evolution of the
diagonal part of the one-body density in r-space. At initial time, the
one-body density has been imposed to be the same for all calculations. We
see that usual TDHF (dashed line) calculation is a good approximation of the
exact one-body dynamics (circles) for time smaller than $100$ $fm/c$.
However, for longer time, we observe strong differences between TDHF
prediction and the expected result. On the same graphic, the equivalent
evolution is reported for ETDHF simulations (solid curve). The ETDHF
prediction of the one-body density operator follow closely the exact one
even for long time evolution.

\begin{figure}[tbph]
\begin{center}
\includegraphics*[height=10cm,width=14cm]{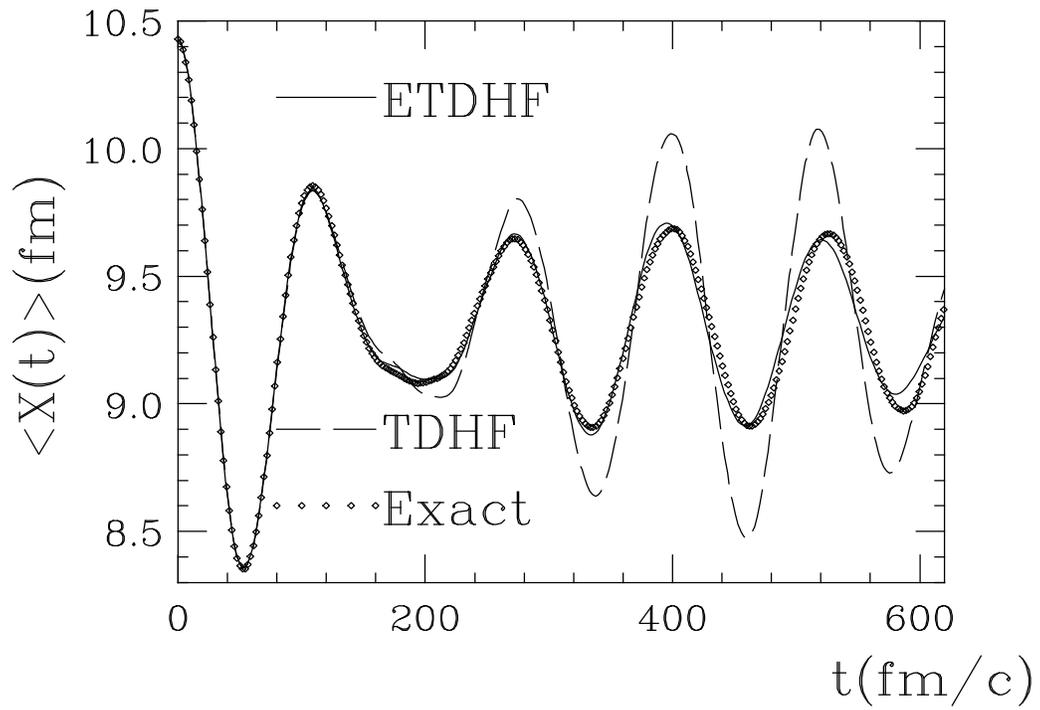}
\end{center}
\caption{Center of mass motion ($<X>$)X of the two-particle in interaction.
The three different simulations are represented: Exact (small circle), TDHF
(dashed line) and ETDHF (solid line).}
\label{fig:6}
\end{figure}

Since the one-body density matrices govern the evolution of all one-body
observables, it means that the extension of mean-field enable us to predict
one-body dynamics with a good accuracy. Indeed, if we look to one particular
one-body observable, for example the center of mass motion (denoted $%
\left\langle X\right\rangle $, fig. \ref{fig:6}), we observe a very good
agreement between exact calculation and ETDHF prediction whereas TDHF fails
to reproduce the long-time dynamics.

\subsubsection{Variation of occupation numbers}

\begin{figure}[tbph]
\begin{center}
\includegraphics*[height=10cm,width=12cm]{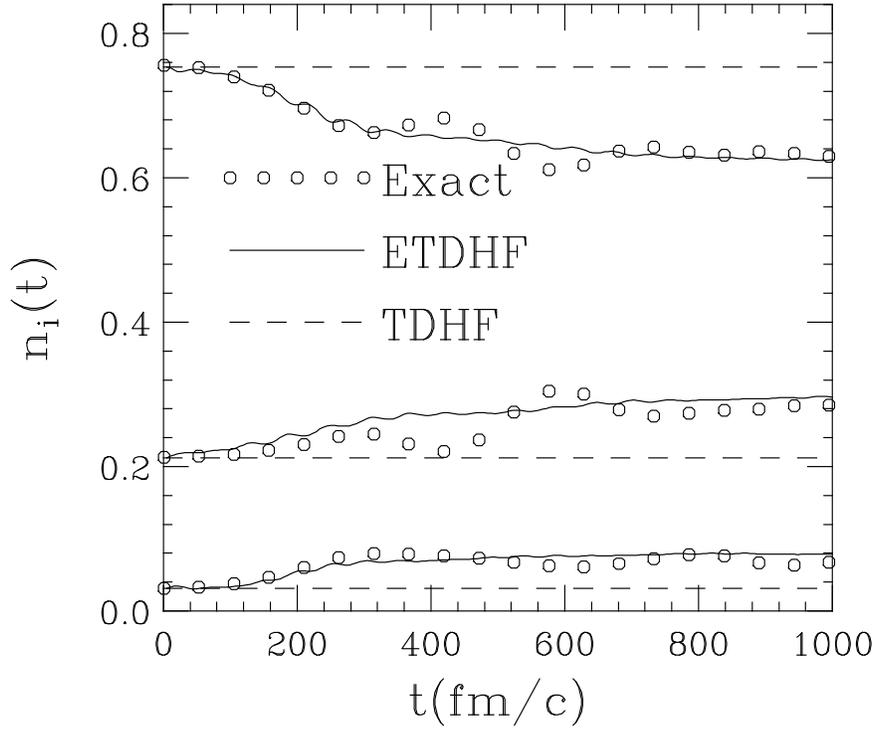}
\end{center}
\caption{Occupation numbers evolution. Circle represents the exact evolution
of occupation numbers and dashed line represents evolution predicted by
ETDHF. The TDHF simulation (not represented) predicts fixed occupation
numbers.}
\label{fig:7}
\end{figure}

In fig. \ref{fig:7}, we have plotted the variation of occupation numbers
(i.e. the eigenvalues of $\hat{\rho}$) in the exact (circles) and ETDHF
treatment (dashed line). Due to the presence of a residual two-body
interaction, we observe a reorganization of occupation numbers in the exact
evolution. The numerical procedure we have developed in order to extend the
mean-field dynamics is able to reproduce this complicated behavior. This
demonstrate in particular that two-body correlation dynamics is properly
taken into account through the method we have used. It is important to note
that reorganization of occupation numbers is accompanied by an evolution of
the relevant single-particle states and that no bias exists in our approach,
in the choice of this relevant states. This is one of the major improvement
contained in our procedure.

\subsubsection{Conclusion and outlook}

In this article, we have discussed the possibility of extending mean-field
in quantum dynamics in order to include two-body correlations. Such theory,
is numerically much more complicated to apply than usual TDHF simulations.
As a result, only few applications have been carried out up to now. These
applications often rely on important simplifications made in the collision
term. We propose here, a procedure to solve numerically this problem. In our
approach, no additional bias is introduced by the simulation and the
collision effect is treated in all its generality. The method is then tested
on a simple model that could be exactly integrated numerically. This model
of two particle interacting one with the other already contains many facet
of many-body problems. In particular, the two-body interaction leads to a
strong reorganization of occupation numbers during the dynamics and affect
considerably the long-time evolution. As expected, usual mean-field dynamics
is not able to account accurately for this complicated behavior. On
opposite, the inclusion of two-body correlation dynamics into the mean
evolution improve considerably the prediction of one-body observables.

The ETDHF seems to be a very promising tool for the description of nuclei
under extreme conditions. In forthcoming work, we will apply the numerical
procedure to realistic nuclei. The inclusion of two-body correlations will
considerably improve our knowledge about the nuclear dynamics. In
particular, this many-body theory includes the possibility for collective
motion to dissipate energy through an irreversible flow to single-particle
degrees of freedom. This will certainly ameliorate the prediction of damping
of giant resonances. On the other hand, this extended theory is needed for
the understanding of thermalization of nuclei.

\section{APPENDIX A: Determination of the instantaneous basis.}

The instantaneous basis is constructed in order to be a good approximation
of part of the eigenvectors of the mean-field Hamiltonian $h$ at time $%
t+\Delta t$. Noting $\left| \xi _\alpha \right\rangle $ this eigenvectors,
we have

\begin{eqnarray}
h\left[ \rho (t+\Delta t)\right] \left| \xi _\alpha \right\rangle
=\varepsilon _\alpha \left| \xi _\alpha \right\rangle
\end{eqnarray}

Determining all $\left| \xi _\alpha \right\rangle $ is not possible in
general since it imply the inversion of huge matrices. However, the $\left|
\xi _\alpha \right\rangle $'s form a complete basis of the single particle
space. In particular, Hartree-Fock states could be expressed in this basis 
\begin{eqnarray}
\left| \Phi _i\right\rangle =\sum_\alpha {C_\alpha ^i\left| \xi _\alpha
\right\rangle }
\end{eqnarray}
Using the imaginary time method, consists in applying the operator $\exp {%
\left( -\beta h\right) }$ to the state: 
\[
\left| \Phi _i^{\prime }\right\rangle =\exp {\left( -\beta h\right) }\left|
\Phi _i\right\rangle 
\]
where $\beta $ is a real number. The interest of such operator leads in the
following: 
\begin{eqnarray}
\left( \exp {\left( -\beta h\right) }\right) ^n\left| \Phi _i\right\rangle
=\exp \left( -\beta n\varepsilon _0\right) \sum_{-\alpha }{C_\alpha ^i\exp
\left( -\beta n\left( \varepsilon _\alpha -\varepsilon _0\right) \right) }%
\left| {\xi _\alpha }\right\rangle
\end{eqnarray}
which shows that 
\begin{equation}
\left( \exp {\left( -\beta h\right) }\right) ^n\left| \Phi _i\right\rangle 
\stackrel{n\rightarrow \infty }{\longmapsto }\left| \xi _0\right\rangle
\end{equation}
We thus see that, application of the imaginary-time operator to any state
converge towards the lowest state in energy. More generally, this operator
remove the space pertained by states with high energy on profit of low
energy states. Using this property, new states are constructed with
following steps:

\begin{itemize}
\item  {\bf Application of $\exp {\left( -\beta h\right) }$}: from each
Hartree-Fock state, a series of state is constructed\footnote{%
Note that, the use of a field $h^{\prime }=h+\delta h$ where $\delta h$ is a
small external stochastic field improve this method.}:

\begin{eqnarray}
\left\{ 
\begin{array}{cl}
\left| \psi _i^{(0)}\right\rangle  & =\left| \Phi _i\right\rangle  \\ 
&  \\ 
\left| \psi _i^{(n)}\right\rangle  & ={\cal N}^{(n)}\left\{ \exp {\left(
-\beta h\right) }-\left\langle \psi _i^{n-1}\right| \exp {\left( -\beta
h\right) }\left| \psi _i^{n-1}\right\rangle \right\} \left| \psi
_i^{(n-1)}\right\rangle 
\end{array}
\right. 
\end{eqnarray}

The number of states per Hartree-Fock states will determine the precision of
the evaluation of the collision term. In practice, we observe that only few
extra states are necessary to be included between two times separated by $%
\Delta t$.

\item  All states are grouped and orthonormalized. Note that the
Hartree-Fock basis remains unchanged.

\item  Instead of considering directly this basis, we diagonalize the
projected Hamiltonian: 
\begin{eqnarray}
(1-\hat{P})h\left[ \rho \right] (1-\hat{P})\left| p_j\right\rangle
=\varepsilon \left| p_j\right\rangle 
\end{eqnarray}
with 
\begin{eqnarray}
\hat{P}=\sum_i{\left| \Phi _i\right\rangle \left\langle \Phi _i\right| }
\end{eqnarray}

\item  Finally states with energy $\varepsilon $ greater than the energy $%
\varepsilon _{Max}$ are rejected. Note that, all created new states belongs
to the kernel of $\hat{\rho}(t+\Delta t)$ which will simplify considerably
the expression of the collision term.
\end{itemize}

\section{APPENDIX B: Expression of transition elements.}

In our instantaneous basis, the density is diagonal, the coefficients of eq. 
\ref{F__} reads 
\begin{eqnarray}
F_{\lambda ,\lambda ^{\prime }} &=&\sum_{\alpha ,\beta ,\delta }{%
\left\langle \lambda \delta |V_{12}|\alpha \beta \right\rangle _A|_t} 
\nonumber \\
&&  \nonumber \\
&&\int_{-\infty }^t{dt^{\prime }}\left( n_{\lambda ^{\prime }}n_\delta
(1-n_\alpha )(1-n_\beta )-n_\alpha n_\beta (1-n_{\lambda ^{\prime
}})(1-n_\delta )\right)  \nonumber \\
&&  \nonumber \\
&&\left\langle \alpha \beta |V_{12}|\lambda ^{\prime }\delta \right\rangle
_A|_{t^{\prime }}  \nonumber
\end{eqnarray}
In order to express the integral in time, we applied a backward mean-field
evolution (where we have neglect the correlation part). In this case,
occupation numbers are constant, however, a finite life-time $\tau $
(equivalent to the Landau beating time) has been added in order to properly
calculate the integral (we have fixed $\tau $ in order to not biased the
integral value).

\begin{itemize}
\item  {\bf Diagonal elements:} The equation of motion is written as
equation (\ref{master}) where 
\begin{eqnarray}
{\cal W}^{-} &=&\frac 4{\hbar ^2}\sum_{\alpha ,\beta ,\delta }{n_\alpha
n_\beta (1-n_\delta )}{\it Real}\left\{ {}\right. \left\langle \lambda
\delta |V_{12}|\alpha \beta \right\rangle _A|_t  \nonumber \\
&&\left. \int_{-\infty }^t{dt^{\prime }}\exp \left( \frac{t^{\prime }-t}\tau
\right) \left\langle \alpha \beta |V_{12}|\lambda \delta \right\rangle
_A|_{t^{\prime }}\right\}   \nonumber \\
{\cal W}^{+} &=&\frac 4{\hbar ^2}\sum_{\alpha ,\beta ,\delta }{(1-n_\alpha
)(1-n_\beta )n_\delta }{\it Real}\left\{ {}\right. \left\langle \lambda
\delta |V_{12}|\alpha \beta \right\rangle _A|_t  \nonumber \\
&&\left. \int_{-\infty }^t{dt^{\prime }}\exp \left( \frac{t^{\prime }-t}\tau
\right) \left\langle \alpha \beta |V_{12}|\lambda \delta \right\rangle
_A|_{t^{\prime }}\right\}   \nonumber
\end{eqnarray}
We have checked that the lost term and the gain term are almost constant
during each time interval $\Delta t$

\item  {\bf Non-diagonal elements:} The non-diagonal elements are necessary
in order to express the mixing of states. The integrated effect of collision
is simply taken as 
\begin{eqnarray}
\left\langle \lambda \right| \Delta \rho \left| \lambda ^{\prime
}\right\rangle =\frac{-\Delta t}{\hbar ^2}\left( F_{\lambda ,\lambda
^{\prime }}+F_{\lambda ^{\prime },\lambda }^{*}\right) 
\end{eqnarray}
\end{itemize}

\end{document}